\documentclass[twocolumn,showpacs]{revtex4}

\usepackage[T1]{fontenc}
\usepackage[cp1250]{inputenc}
\usepackage{graphicx}
\usepackage{dcolumn}
\usepackage{bm}
\usepackage{ulem}

\newcommand{\vect}[1]{{\boldsymbol{\mathbf{#1}}}}

\begin{document}

\title{Nonlinear magneto-optical rotation with modulated light in tilted magnetic fields}

\author{S.~Pustelny}
\affiliation{Centrum Bada\'n Magnetooptycznych, M.~Smoluchowski
Institute of Physics, Jagiellonian University, Reymonta 4, 30-059
Krak\'ow, Poland}
\author{S. M.~Rochester}
\affiliation{Department of Physics, University of California at
Berkeley, Berkeley, CA 94720-7300, USA}
\author{D.~F.~Jackson~Kimball}
\affiliation{Department of Physics, California State University --
East Bay, 25800 Carlos Bee Blvd., Hayward, CA 94542, USA}
\author{V.~V.~Yashchuk}
\affiliation{Advanced Light Source Division, Lawrence Berkeley
National Laboratory, Berkeley CA 94720, USA}
\author{D.~Budker}
\affiliation{Department of Physics, University of California at
Berkeley, Berkeley, CA 94720-7300, USA} \affiliation{Nuclear Science
Division, Lawrence Berkeley National Laboratory, Berkeley CA 94720,
USA}
\author{W.~Gawlik}
\affiliation{Centrum Bada\'n Magnetooptycznych, M.~Smoluchowski
Institute of Physics, Jagiellonian University, Reymonta 4, 30-059
Krak\'ow, Poland}

\date{\today}

\begin{abstract}
Larmor precession of laser-polarized atoms contained in
anti-relaxation-coated cells, detected via nonlinear magneto-optical
rotation (NMOR) is a promising technique for a new generation of
ultra-sensitive atomic magnetometers.  For magnetic fields directed
along the light propagation direction, resonances in NMOR appear
when linearly polarized light is frequency- or amplitude-modulated
at twice the Larmor frequency.  Because the frequency of these
resonances depends on the magnitude but not the direction of the
field, they are useful for scalar magnetometry.  New NMOR resonances
at the Larmor frequency appear when the magnetic field is tilted
away from the light propagation direction in the plane defined by
the light propagation and polarization vectors.  These new
resonances, studied both experimentally and with a density matrix
calculation in the present work, offer a convenient method for
NMOR-based vector magnetometry.
\end{abstract}

 \maketitle

When linearly polarized light, frequency- or amplitude-modulated at
$\Omega_m$, resonantly interacts with an atomic vapor in the
presence of a magnetic field, the polarization of the light can be
observed to rotate synchronously with the modulation. This effect is
known as nonlinear magneto-optical rotation with frequency-modulated
light (FM NMOR) \cite{FMNMOR,FMNMORReview,Malakyan2004} or
amplitude-modulated light (AMOR) \cite{AMOR,Balabas2006}. It occurs
when $\Omega_m$ is a subharmonic of the quantum beat frequency; the
quantum beat frequency is at the first or second harmonic of the
Larmor frequency $\Omega_L$ for the lowest-order effect discussed
here. Higher-order effects involve quantum beat frequencies at other
multiples of $\Omega_L$ \cite{Selective}. The width of the resonance
between $\Omega_m$ and $\Omega_L$ is given by the relaxation rate of
the atomic ground-state coherences. When the atomic vapor is
contained in a paraffin-coated cell, in which ground-state atomic
coherences can survive on the order a second, the widths of the
resonances can be as small as 0.6 Hz \cite{Budker2005}. These
resonances allow extremely precise measurements of the magnetic
field over a wide field range, with magnetometric sensitivities
exceeding $10^{-11}~\mbox{G}/\sqrt{\mbox{Hz}}$ for low fields
(competitive with the best atomic magnetometers, see for example
Ref. \cite{SERF}, as well as SQUID magnetometers \cite{SQUID}) and
reaching $4\times 10^{-10}~\mbox{G}/\sqrt{\mbox{Hz}}$ for higher
fields (up to 1 G) \cite{EarthField}. The FM NMOR method used in the
present work, as well as AMOR, can be applied to studies of nuclear
magnetic resonance and magnetic-resonance imaging
\cite{Yas04,Xu2006}, measurements of geophysical fields
\cite{EarthField}, and tests of fundamental symmetries
\cite{Kimball2001AIP}. The same approach can also be used in
construction of chip-scale atomic magnetometers \cite{Schwindt2005}.

Although the magnetometric method based on FM NMOR enables sensitive
measurements of the magnetic field, the measurements are scalar,
i.e., the position of a given resonance depends only on the
magnitude, and not the direction, of the magnetic field. However,
the relative magnitudes of the FM NMOR resonances can depend on the
magnetic field direction. Thus, a detailed analysis of the FM NMOR
signal could give some information about the direction of the
magnetic field. Towards this end, we study here the dependence of
the FM NMOR signal on the magnetic field direction.

In the Faraday geometry, in which the magnetic field is along the
light propagation direction, the main resonance occurs at
$\Omega_m=2\Omega_L$, because of the symmetry of the optically
pumped state, as discussed below. We find that when the magnetic
field direction is tilted in the plane perpendicular to the light
polarization axis, only this resonance at $2\Omega_L$ is observed,
with its amplitude depending on the tilt angle. However, when the
magnetic field is tilted toward the light polarization axis, a new
resonance appears at $\Omega_L$; the relative magnitudes of the
resonances at $\Omega_L$ and $2\Omega_L$ depend on the tilt angle.

\begin{figure}[htb!]
    \includegraphics{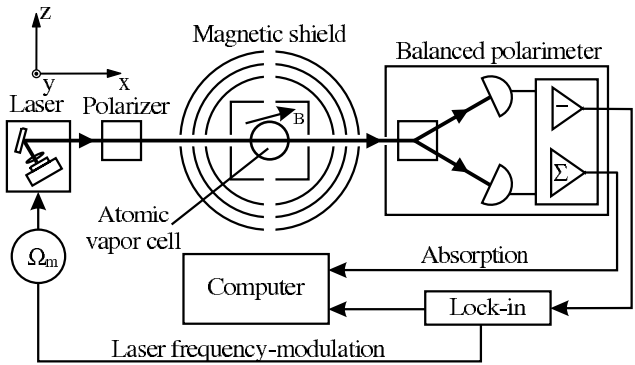}
    \caption{Experimental setup. The magnetic field coils mounted inside the
    innermost shielding layer and used for creation of arbitrarily
    oriented magnetic field are not shown.} \label{fig:setup}
\end{figure}

The scheme of the experiment is shown in Fig. \ref{fig:setup}.
An anti-relaxation-coated buffer-gas-free vapor cell, containing
isotopically enriched $^{85}$Rb, was placed within a four-layer
magnetic shield. The magnetic shield provided passive attenuation of
dc magnetic fields by a factor of 10$^6$ \cite{YashchukShield}. A
set of three mutually orthogonal magnetic-field coils placed inside
the innermost layer enabled compensation of the residual average
magnetic field and first-order magnetic field gradients inside the
shield. The coils were also used for generation of an arbitrarily
oriented magnetic field inside the shield. The rubidium atoms
interacted with an $x$-directed, 2 mm diameter laser light beam,
linearly polarized along the $y$-axis. An external cavity diode
laser was tuned to the rubidium D$_1$ line (795 nm) and its central
frequency was stabilized with a dichroic atomic vapor laser lock
\cite{DAVLLWieman,BudkerDAVLL} at the low-frequency wing of the
$F=3\rightarrow F'$ transition. The laser-modulation frequency
ranged from 100 Hz to 400 Hz with 300 MHz (peak to peak) modulation
depth and light power was $\sim$3 $\mu$W. Upon passing through the
cell, the light was analyzed using a balanced polarimeter. The
polarimetric signal was detected with a lock-in amplifier at the
first harmonic of $\Omega_m$ and stored on a computer.

\begin{figure}[htb!]
    \includegraphics{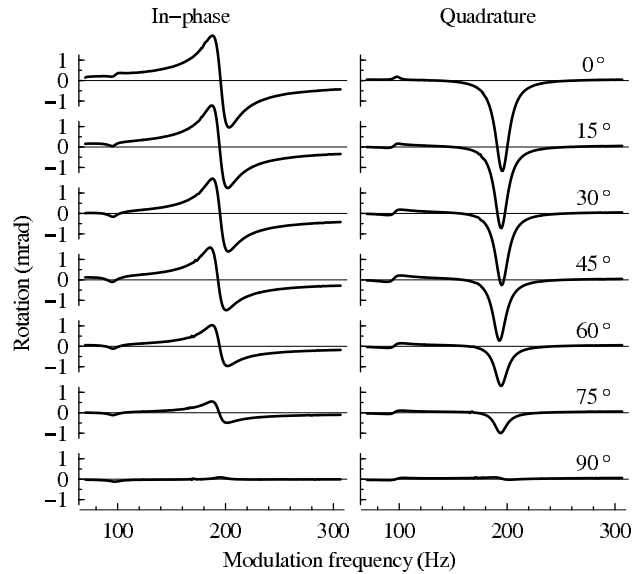}
    \caption{The FM NMOR in-phase and quadrature signals vs.
    modulation frequency recorded for various angles between the
    magnetic field and the light propagation direction in the
    $xz$-plane. A small residual signal observed for 90$\circ$
    is a result of a small $y$-component of the magnetic field.
    } \label{fig:tiltingfieldXZ}
\end{figure}

\begin{figure}[htb!]
   \includegraphics{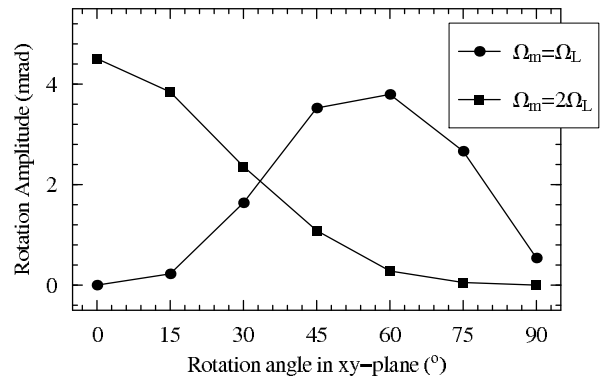}
   \caption{The amplitude of the FM NMOR signals recorded at 2$\Omega_L$
   vs. the tilt angle of the magnetic field in the $xz$-plane. The solid
   line is a cosine fit to the experimental points.}
   \label{fig:AmplitudeVsAngleXZ}
\end{figure}

\begin{figure}[htb!]
  \includegraphics{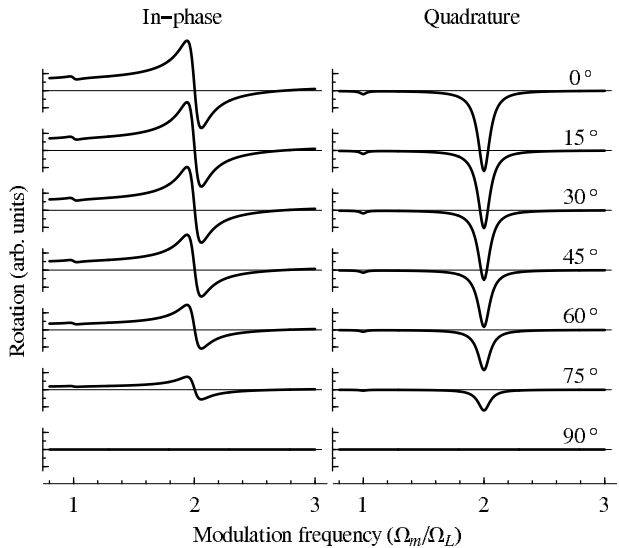}
  \caption{Theoretical calculation of FM NMOR signal vs.\
  modulation frequency for a $F=1\rightarrow F=0$ transition for
  various tilt angles of the magnetic field in the $xz$-plane.}
  \label{fig:TheoryModFreqSpecXZ}
\end{figure}

The FM NMOR signals were studied as the magnetic field direction was
tilted in both the $xz$-plane (perpendicular to the
light-polarization axis) and the $xy$-plane (containing the
light-propagation vector and light-polarization axis). Figure
\ref{fig:tiltingfieldXZ} shows the signals with the magnetic field
$\vect{B}$ in the $xz$-plane at various angles to the light
propagation direction.
The strength of the magnetic field was equal for all the
measurements ($\Omega_L=98.5$ Hz). The main resonance occurs at
$\Omega_m=2\Omega_L$. The small resonance at $\Omega_L$ is a result
of using the modulated pump beam as a probe. For the
frequency-modulated probe the time dependence of the
light-polarization plane is observed, even if the atomic
polarization in the medium is static. This effect has been analyzed
in detail in Ref.\ \cite{Pustelny2006} Sec.\ IV. The amplitudes of
the resonances decrease with increasing angle in the $xz$-plane
(Fig.\ \ref{fig:AmplitudeVsAngleXZ}).
The experimental signals are in qualitative agreement with the
predictions of a density-matrix calculation \footnote{The optical
Bloch equations are solved using the ``matrix-continued-fraction
method'' in the manner of Ref. \cite{Nayak1985}: the density matrix
is expanded in a Fourier series in harmonics of the modulation
frequency, and the resulting matrix recursion relation for the
Fourier coefficients is inverted as a matrix continued fraction,
which is evaluated by assuming a high-harmonic cutoff.} of FM NMOR
for a model $F=1\rightarrow F'=0$ transition (Fig.\
\ref{fig:TheoryModFreqSpecXZ}).

\begin{figure}[htb!]
    \includegraphics{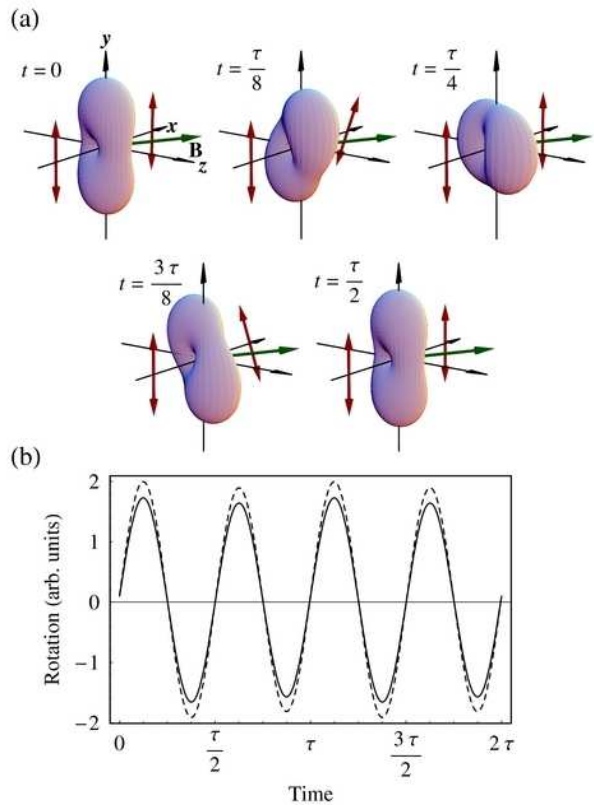}
    \caption{Illustration of the time dependence of FM NMOR for
    $\vect{B}$ in the $xz$ plane at $30^\circ$ to the light
    propagation direction $\hat{\vect{x}}$, using results of the
    calculation for a $F=1\rightarrow F'=0$ transition. (a) Angular
    momentum probability surfaces depicting the dynamic part of the
    atomic alignment at various times during the Larmor period
    $\tau$. The alignment returns to its original position after
    time $\tau/2$. The double-headed arrows represent the light
    polarization before and after the medium. (b) Optical rotation
    as a function of time for the case shown in part (a) (solid
    line), and for $\vect{B}$ along $\hat{\vect{x}}$ (dashed line).
    The dominant signal has period $\tau/2$ (frequency $2\Omega_L$),
    corresponding to the periodicity of the polarization state shown
    in part (a). The effect of using the modulated pump as a probe
    beam can be seen here as a small additional modulation with
    period $\tau$.} \label{fig:AlignmentRotationInXZ}
\end{figure}

The observed dependence on the magnetic-field direction can be
understood by considering the time evolution of the atomic
polarization. In the absence of a magnetic field, the linearly
polarized light creates atomic alignment (polarization with a
preferred axis, but no preferred direction) along the $y$-axis. This
alignment can be visualized with a method described in, for example,
Ref.\ \cite{SimonVisualization}: a surface is plotted whose radius
in a given direction is equal to the probability of finding the
maximum projection of angular momentum along that direction. When a
magnetic field is applied, the alignment precesses around the
magnetic-field direction at the Larmor frequency [Fig.\
\ref{fig:AlignmentRotationInXZ}(a)].
Since in this case $\vect{B}$ is perpendicular to the alignment
axis, the polarization returns to its original state after a
$180^\circ$ rotation, i.e., in half a Larmor period \footnote{In
this work the light power was sufficiently low to exclude effects
such as alignment-to-orientation conversion \cite{AOC} that cause
more complicated polarization evolution.}. Thus optical rotation,
induced by the rotating linear dichroism, is periodic at twice the
Larmor frequency [Fig.\ \ref{fig:AlignmentRotationInXZ}(b)]. Because
only linear dichroism transverse to the light propagation direction
(i.e., in the $yz$ plane) can produce rotation, the amplitude of the
rotation decreases as the cosine of the angle between $\vect{B}$ and
the light propagation direction, as seen experimentally in Fig.\
\ref{fig:AmplitudeVsAngleXZ}.

\begin{figure}[htb!]
    \includegraphics{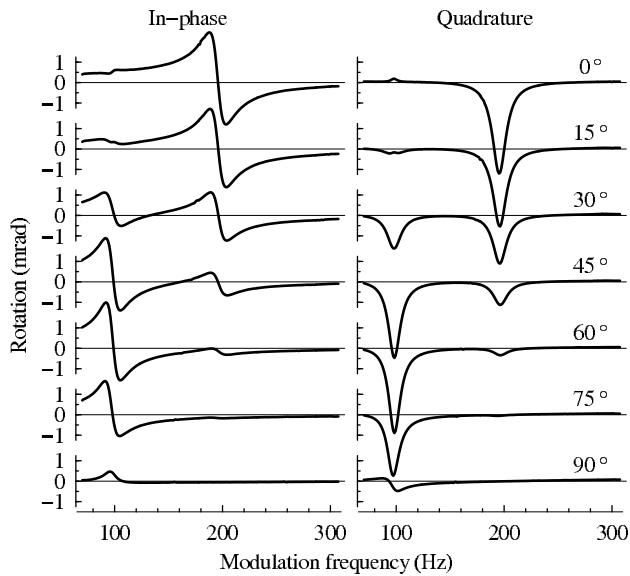}
    \caption{The FM NMOR signal vs. modulation frequency $\Omega_m$
    recorded for different tilt angles of the magnetic field
    in the $xy$-plane. We proved with theoretical
    simulations that the small signal observed at $\Omega_L$ for the
    magnetic field oriented along $y$-axis is due to the misalignment between
    the magnetic field and the light polarization direction.}
    \label{fig:tiltingfieldXY}
\end{figure}

\begin{figure}[htb!]
  \includegraphics{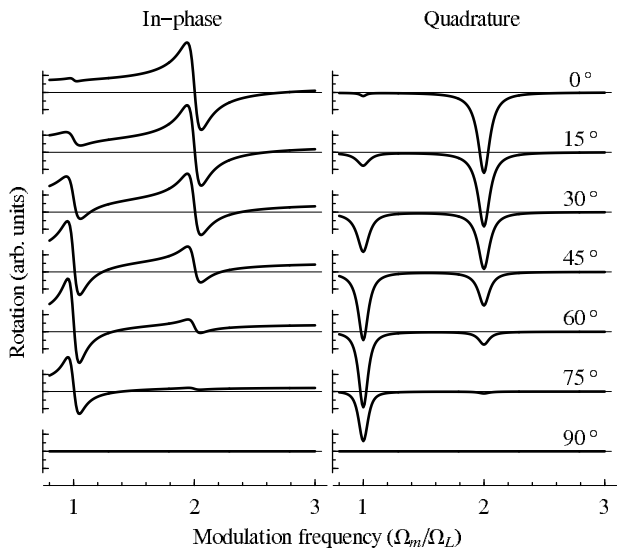}
  \caption{Theoretical calculation of FM NMOR signal vs.\
  modulation frequency for a $F=1\rightarrow F=0$ transition for
  various tilt angles of the magnetic field in the $xy$-plane.}
  \label{fig:TheoryModFreqSpecXY}
\end{figure}

\begin{figure}[htb!]
   \includegraphics{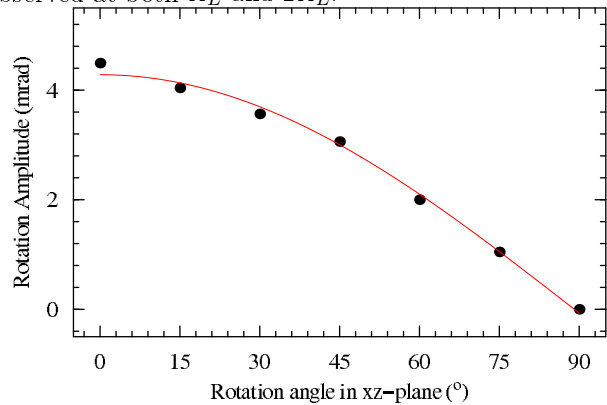}
   \caption{The amplitude of the FM NMOR signals recorded at $\Omega_L$
   and 2$\Omega_L$ vs. the tilt angle of the magnetic field in $xy$-plane.}
   \label{fig:AmplitudeVsAngleXY}
\end{figure}

Figure \ref{fig:tiltingfieldXY} shows the FM NMOR signals for the
magnetic field tilted in the $xy$ plane at various angles to the
light propagation direction; corresponding theoretical results for a
$F=1\rightarrow F'=0$ transition are shown in Fig.\
\ref{fig:TheoryModFreqSpecXY}.
When the magnetic field is parallel to the light propagation
direction a strong FM NMOR resonance is observed only at
$\Omega_m=2\Omega_L$. However, when the magnetic field is tilted
toward $y$-axis an additional resonance appears at
$\Omega_m=\Omega_L$. The amplitude of this resonance becomes more
pronounced with increasing the angle, while the amplitude of the
resonance at 2$\Omega_L$ decreases (Fig.\
\ref{fig:AmplitudeVsAngleXY}).
However, when the magnetic field is tilted by more than 60$^\circ$
the amplitude of the resonance recorded at $\Omega_L$ also starts to
decrease and reaches zero when the magnetic field is directed along
the $y$-axis.

\begin{figure}[htb!]
    \includegraphics{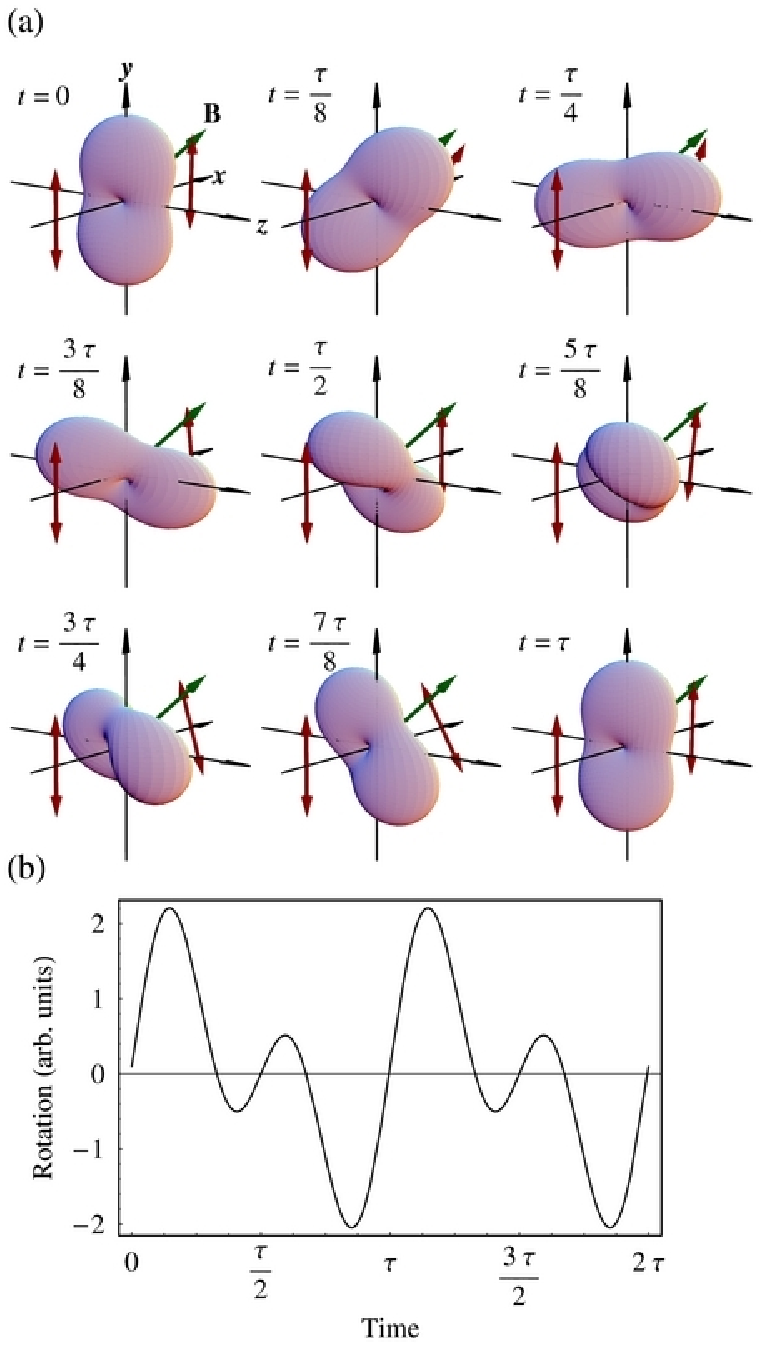}
    \caption{As Fig.\ \ref{fig:AlignmentRotationInXZ}, but with
    $\vect{B}$ in the $xy$ plane at $17.5^\circ$ to the light
    propagation direction $\hat{\vect{x}}$. (a) The alignment axis
    in this case is not perpendicular to the magnetic field, so it
    takes a full Larmor period $\tau$ to return to its original
    state. (b) The optical rotation. The optical anisotropy in the
    $yz$-plane has components at $\Omega_L$ and $2\Omega_L$.}
    \label{fig:AlignmentRotationInXY}
\end{figure}

Once again, these signals can be explained by considering the
time-evolution of atomic polarization [Fig.\
\ref{fig:AlignmentRotationInXY}(a)].
Because the magnetic field is no longer perpendicular to the
alignment axis, the polarization takes a full Larmor period to
return to its original state. The anisotropy of the medium in the
$yz$-plane is modulated at two frequencies: $\Omega_L$ and
$2\Omega_L$, as reflected in the plot of time-dependent optical
rotation [Fig.\ \ref{fig:AlignmentRotationInXY}(b)]. The larger the
tilt angle between $\vect{B}$ and the light propagation direction,
the bigger the difference between the polarization states at the
beginning and middle of the Larmor periods, increasing the signal at
$\Omega_L$. However, for large angles, the magnetic field direction
is nearly along the alignment axis, reducing the effect of the field
on the polarization and thus reducing the optical rotation.

The data presented in Figs. \ref{fig:tiltingfieldXZ} and
\ref{fig:tiltingfieldXY} can be also understood in a language of
atomic coherences. When the magnetic field is in the $xz$-plane, the
light polarization axis is perpendicular to the magnetic field.
Thus, with the quantization axis along the magnetic field, the light
can only create coherence between the $m=1$ and $m=-1$ Zeeman
sublevels. The frequency splitting between these sublevels is
2$\Omega_L$, so the resonance is observed at this frequency.
However, when the magnetic field is tilted in the $xy$-plane, the
light is a linear superposition of polarizations parallel and
perpendicular to the magnetic field. In this case, the light can
create coherences between sublevels with $\Delta m=1$ and $\Delta
m=2$, so resonances are observed at both $\Omega_L$ and 2$\Omega_L$.

In conclusion, we have studied resonances in nonlinear
magneto-optical rotation with frequency-modulated light (FM NMOR)
when the magnetic field is tilted at angles away from the direction
of light propagation.  When the magnetic field is tilted in the
plane defined by the light polarization and light propagation
vectors, a new FM NMOR resonance appears at the Larmor frequency.
The amplitude of the resonance depends on the tilt angle and thus
can be useful for vector magnetometry or for aligning a scalar
magnetometer to reduce heading errors.  When the magnetic field is
tilted in the orthogonal plane, no new resonance appears at the
Larmor frequency. For tilt angles in both planes, the amplitude of
the FM NMOR resonance at twice the Larmor frequency decreases with
increasing tilt angle. The effects have been modeled with density
matrix calculations, yielding good agreement with the experimental
observations, and can be understood through visualization
\cite{SimonVisualization} of the time-dependence of the atomic
polarization moments.

\begin{acknowledgments}
The authors would like to acknowledge M. Auzinsh, M. P. Ledbetter,
A. Cingoz, and J. Higbie for fruitful discussions. This work is
supported by DOD MURI grant \# N-00014-05-1-0406, KBN grant \# 1
P03B 102 30, and a NSF US-Poland collaboration grant. One of the
author (S.P.) is a scholar of the project co-financed from the
European Social Fund.
\end{acknowledgments}

\end{document}